\documentclass[12pt]{article}

\usepackage{scicite}
\usepackage{times}
\usepackage{graphicx}

\newenvironment{sciabstract}{%
\begin{quote} \bf}
{\end{quote}}

\title{Plasma-induced surface cooling}

\author
{John A. Tomko,$^{1}$ Michael J. Johnson,$^{2}$ David R. Boris,$^{3}$\\
Tzvetelina B. Petrova,$^{3}$ Scott G. Walton, $^{3\ast}$ Patrick E. Hopkins, $^{1,4,5\ast}$\\
\\
\normalsize{$^{1}$Department of Mechanical and Aerospace Engineering, University of Virginia,}\\
\normalsize{Charlottesville, VA 22904, USA}\\
\normalsize{$^{2}$Syntek Technologies}\\
\normalsize{Fairfax, VA 22031, USA}\\
\normalsize{$^{3}$Plasma Physics Division, Naval Research Laboratory}\\
\normalsize{Washington DC 20375, USA}\\
\normalsize{$^{4}$Department of Materials Science and Engineering, University of Virginia,}\\
\normalsize{Charlottesville, VA 22904, USA}\\
\normalsize{$^{5}$Department of Physics, University of Virginia,}\\
\normalsize{Charlottesville, VA 22904, USA}\\
\normalsize{$^\ast$ To whom correspondence should be addressed;}\\
\normalsize{E-mail: scott.walton@nrl.navy.mil; phopkins@virginia.edu}
}

\date{}

\begin{document}

\baselineskip24pt

\maketitle

\begin{sciabstract}
Plasmas have long been used for the synthesis \cite{Dou2018} and manipulation \cite{Donnelly2013,Chu2002,Penkov} of materials because of their unique ability to deliver both energy and chemically-active species to the surface of materials - an attribute that separates them from other approaches to materials processing. This energy flux serves to drive the surfaces out of thermal equilibrium with the bulk material, thus enabling local physicochemical processes that can be harnessed to remove (etch) substrate material, deposit different material, or chemically modify the surface. However, to-date, there have been no reports on the direct measurement of the \textit{localized, transient response} of a material surface subjected to a plasma energy flux. In other words, a direct \textit{in-situ} measure of the transient thermal response of a material subjected to a plasma flux is lacking. Here we show that, despite this massive incident flux of energetic species, plasmas can induce transient cooling of a material surface. Using time-resolved optical thermometry \textit{in-situ} with this plasma excitation, we reveal the novel underlying physics that drive this `plasma cooling' that is driven by the diverse chemical and energetic species that comprise this fourth state of matter. We show that the photons and massive particles in the plasma impart energy to different chemical species on a surface, leading to local and temporally changing temperatures that lead to both increases and \textit{decreases} in temperature on the surface of the sample, even though energy is being imparted to the material. This balance comes from the interplay between chemical reactions, momentum transfer, and energy exchange which occur on different time scales over the course of picoseconds to milliseconds. Thus, we show that through energetically exciting a material with a plasma, we can induce $cooling$, which can lead to revolutionary advances in refrigeration and thermal mitigation with this new process that is not inhibited by the same limitations in the current state-of-the-art systems.

\end{sciabstract}

\paragraph*{Introduction.}  Plasmas have long been used for the synthesis \cite{Dou2018} and manipulation \cite{Donnelly2013,Chu2002,Penkov} of materials because of their unique ability to deliver both energy and chemically-active species to the surface of materials (Fig.~1) - an attribute that separates them from other approaches to materials processing. Indeed, the energy flux serves to drive the surfaces out of thermal equilibrium with the bulk material, thus enabling local physicochemical processes that can be harnessed to remove (etch) substrate material, deposit different material, or chemically modify the surface. Aside from intentional material modifications, understanding energy delivery at the plasma-surface interface is critical for an array of technologies such as nuclear fusion, where plasma-facing materials must meet complex, yet strict, requirements to avoid degradation from the aforementioned energetic processes \cite{Linsmeier2017}. While the benefits or detriments of energy delivery is commonly associated with an increase in temperature, the temperature, is in fact, the net result of the difference between energy delivered to and released from the surface. This can be understood by considering the power balance at the surface \cite{Kersten2001},
\begin{equation}
P_{in} - P_{out} = P_{heat}
\end{equation}
where the power delivered to ($P_{in}$) and released from ($P_{out}$) the surface is determined by the flux of energetic particles and radiation arriving at and leaving the surface, along with endothermic and exothermic reactions occurring at the plasma-surface interface. The difference between $P_{in}$ and $P_{out}$ is absorbed by the material ($P_{heat}$), with a temperature determined by the thermophysical properties of the material. Of course, this power balance does not dictate that the energy delivered to the surface must exceed that released from the surface. The complex array of incident plasma species and chemical reactions at the surface could, in theory, enable local temperatures to both increase or decrease during plasma irradiation. This potential for \textit{plasma-induced cooling}, or the decrease in the temperature of a surface during plasma irradiation, could provide novel avenues for structure and device cooling, refrigeration, and temperature-controlled material processing. 

Our current understanding of energy delivery from a plasma to a material surface and its response is guided using a variety of ancillary plasma diagnostics \cite{Hutchinson1987}, steady-state temperature measurements \cite{Takizawa2004,Staack2004}, models \cite{Graves2009,Warrier2004}, and post-treatment \textit{ex-situ} surface characterization to `re-construct' energy deposition and absorption \cite{Kersten2000,Kersten2001,Knoll2014}. More recently, \textit{in-situ} materials characterization techniques have been developed that allow for real time or quasi-real time analysis \cite{Walton2018,Shinohara2003}. While certainly of value, none of these approaches provide a direct measure of the  response associated with the flux of species at the surface required to separate the localized and transient energy transport mechanisms from the spatially and temporally averaged net power transfer and temperature rise. 

In this work, we experimentally demonstrate the ability of an incident plasma to cool the surface of a material.  While seemingly counterintuitive that an incident flux of energy would lead to a decrease in surface temperature, this plasma cooling is enabled by exposing a surface to a pulsed plasma, which allows the broad range of different energetic processes associated with plasma exposure to be parsed in time. This cooling is then measured through time-resolved, relative temperature changes in the plasma-exposed material with nanosecond resolution.

Specifically, we expose an 80 nm gold (Au) film supported by a sapphire substrate to a pulsed, atmospheric plasma jet (Fig.~1) and simultaneously measure the reflectance of a continuous wave laser from the Au surface at the point of contact. A simplified schematic of our experimental configuration is shown in Fig.~S1.  For the operating conditions in this work, there are negligible laser-plasma interactions, and the reflected beam is not affected by any direct interactions. Rather, we rely on the strong thermoreflectance coefficient of Au at visible wavelengths\cite{Wilson2012,Radue2018} to directly measure the plasma-induced temperature change on the Au surface by means of lock-in detection at the plasma jet repetition frequency, to obtain nanosecond time resolution. While we do not observe any changes in the $static$ reflectivity of the Au surface, plasma effects are further isolated by measuring the differential reflectivity, which is the change in reflectivity of the Au surface relative to the reflectance when no plasma is present. 

The use of a thin Au film supported by an insulting substrate satisfies two criteria that are critical to understanding energy transfer at the plasma-surface interface. First, as Au is a noble metal and chemically inert, there are minimal surface reactions (e.g., surface oxidation) that would significantly alter the surface and distort the interpretation of energy deposition mechanisms. Second, because charge transfer and hot-electron effects occur on much faster time scales than investigated in this work, the insulating substrate ensures that charged species and electronically-driven energy transfer from the plasma to the metal surface remain localized to the surface of the Au film. This critical aspect ensures that the measured surface temperature is indicative of only the plasma-Au energy transfer and subsequent thermal diffusion rather than ballistic mechanisms that traverse deep into the substrate \cite{Tomko2020,Giri2015b}.

\paragraph*{Results.} In considering plasma interactions, there is an incident flux of various species including charged particles, photons, as well as excited and reactive neutrals (Fig.~1), which deliver energy to the plasma-exposed material \cite{Kersten2001}. In this work, we employ a pulsed, plasma jet produced in helium, which interacts with a gold surface located some distance from the jet (See the supplemental information for more detail). While these sources produces species common to low temperature, non-equilibrium plasmas, the are several unique attributes worth noting that are relevant to the work here. When the high-voltage pulse is applied to the active electrode within the jet body, a high intensity, ionization wave – or `streamer' - is ejected from the jet nozzle and guided by the helium flow through the ambient \cite{Boeuf2013}. The streamer velocity is significantly faster than the helium gas flow and quickly terminates at the gold surface located downstream from the jet. The plasma channel left in the wake of the streamer then persists for as long as the voltage is applied to the electrode. The easy to ionize helium column remains remarkably undiluted in the center of the column \cite{VanDoremaele2018,Johnson2021}.  As such, species production extends over a fixed volume, but the intensity varies in time. The flux of charged particles - ions and electrons - can be measured as a net surface current, as shown in Fig.~2a for our experimental system. In these example data, the plasma jet is produced by applying a high voltage (2000 V) for a duration (pulse width) of 5 $\mu$s at a frequency of 7.8 kHz. The delay in the rise in current is the time required for the streamer to travel the distance between the powered electrode and the surface. Likewise, as the reflectance of Au is linearly proportional to temperature (e.g., $\Delta R_{f} = \beta \Delta T$, where $\beta$ is the thermoreflectance coefficient of Au)\cite{Rosei1972}, we can directly measure the temperature rise induced by the flux of charged particles with our laser probe; an example of our measured thermoreflectance data is shown in Fig.~2b. Note, the thermoreflectance coefficient for this laser wavelength (637 nm) is negative, and thus surface heating corresponds to a decrease in measured reflectance \cite{Wilson2012,Radue2018}. As shown in the Supporting Information (Fig.~S2), the signal is inverted for probe wavelengths corresponding to photon energies exceeding the interband transition threshold of Au ($<$ 520 nm) and the thermoreflectance coefficient becomes a positive value. 

A few salient features can be noted in Fig.~2b. First, at the beginning and end of the plasma pulse, an anomalous decrease in signal magnitude can be observed. This is an artifact in our periodic waveform analyzer and \textit{not} a true change in optical reflectance due to the plasma pulse; these features are present even when the laser is turned off and the photodetector is blocked. Second, there is a simultaneous, rapid increase of both the measured thermoreflectance and surface current about 2 $\mu$s after the rise in voltage. This heating event is followed by a transient decay associated with heat conduction into the substrate, which is governed by the thermal properties of the film and substrate. Interestingly, a peak in reflectivity is observed about 1 $\mu$s after the rise in voltage, and prior to the rapid heating event. As the thermoreflectance coefficient of Au at this laser wavelength (637 nm) is negative, this increase suggests a \textit{reduction} in surface temperature. 

As discussed above, the flux of various charged species incident upon the metal film results in a net current, which leads one to invoke Joule's first law where the power dissipated is proportional to the product of the current and resistance (e.g., $P \propto I^{2}R$). As with other current sources, this relationship explicitly leads to an increase in temperature, even without consideration of particle interactions (e.g., momentum transfer due to the kinetic energy of incoming ions or the charge transfer associated with neutralization of ions). As the measured thermoreflectance is indicative of a change in the gold's temperature, one should expect that the temporal derivative of this reflectivity trends with the current, since temperature is directly related to particle number density/charge, which is the temporal integrand of current (e.g., $R_{f} \propto T \propto Q = \int{i\cdot dt}$). Indeed, as shown in Fig.~2c, the temporal derivative of our thermoreflectance data is in excellent agreement with the measured current. We thus conclude that heating is, as assumed, associated with simple ohmic heating of the metal film. In stark contrast is the strong deviation between current and derivative associated with the observed reduction in temperature prior to heating. This leads to the obvious question: What is the mechanism for this plasma-induced cooling?

\paragraph*{Discussion.}  Although the notion of cooling resulting from an incident energy flux is seemingly counter-intuitive, there are a number of physical processes arising at the plasma-surface interface that could potentially reduce the temperature of a surface. Thermionic emission, for example, has been theoretically devised as a refrigeration method, with potential efficiencies on par with Carnot cycles \cite{Mahan1994}, though it is nearly impossible to experimentally achieve at or near room temperature, with the exception of limited cases in select material systems such as superlattices \cite{Vashaee2004} and 2-D heterostructures \cite{Rosul2019}. Of course, in the case of a low-temperature atmospheric plasma jet as studied in this work, this thermally-driven mechanism of electron cooling cannot occur.

However, there are two additional effects that could also lead to the observed cooling phenomenon. The first, which has been recently shown to lead to a temperature decrease in high-repetition laser ablation\cite{Kerse2016}, is material ejection from the surface. It is well-known that plasma jets lead to modification of a material surface. In fact, when the jet used in this work is operated at higher power (much greater than reported here), erosion of the Au surface is observed. Alternatively, adsorbate desorption could be the underlying mechanism for observed cooling. Despite the inert nature of Au, water will adsorb on Au surfaces even under ultra-high vacuum conditions \cite{Freund1999}, albeit in a weakly bound physisorbed state ($<< 1$ eV). In such a case, many species emanating from the plasma jet would deliver enough energy to liberate water molecules. This process is effectively plasma-induced evaporative cooling of the surface.

The second potential mechanism is similar to the Nottingham effect, whereby cooling results from the loss of electrons via field emission \cite{Fleming1940}. In this work, electrons removed from the surface of the gold during the neutralization of ions as they approach the metal surface \cite{Hagstrum1954} or via photoemission may lead to cooling. 

To elucidate which of these potential mechanisms is responsible for plasma-induced surface cooling, it is important to consider the magnitude of temperature reduction upon cooling. To gain insight to this, we note that during the observed cooling, the peak differential reflectivity is measured to be $\Delta R_{f}/R_{f} \approx 3\times10^{-5}$. Based on an array of previous works \cite{Wilson2012,Radue2018}, the thermoreflectance coefficients of thin Au films are of similar order, $\Delta R_{f}/\Delta T \approx 2-4\times10^{-5}$, indicating the observed cooling is on the order of approximately 1 K. 

If we consider the possibility that this temperature reduction is due to the sublimation of the Au film, then only a monolayer-equivalent number of Au atoms would need to be ejected from the film within the plasma-irradiated region to achieve the observed degree of cooling (see Supporting Information for calculations and further discussion on this topic). Similarly, if we consider that the observed cooling is induced by the removal of adsorbed water, which has a significantly greater heat capacity than that of Au, a similar temperature decrease requires sub-monolayer, or non-uniform distributions, of water to be removed from the surface within the probed volume. This is certainly plausible given our measurements are conducted at standard temperature and pressure (STP) \cite{Freund1999}, where re-adsorption occurs on timescales orders of magnitude shorter than the period between plasma pulses (tens of nanoseconds-to-microseconds for contaminant adsorption, compared to the hundreds of microseconds between each measured plasma pulse). Likewise, a mixture of adsorbed hydrocarbons (e.g., adventitious carbon) and gas would be expected under these conditions. We note that our approximation is likely an overestimation of the average energy per particle; due to finite coverage of the adsorbed layers, the phonon population is restricted for measurements at room temperature and is likely restricted to a classical equipartition limit. Additionally, adsorbate-adsorbate and adsorbate-substrate interactions (e.g., binding energies) will likely play enough of a role that the energy-per-particle is altered. 

We can repeat a similar analysis for the number of electrons that would need to be emitted for a 1 K temperature reduction within the probed volume of the Au; at the time-scales measured here, the electron and phonon temperatures can be considered in equilibrium, and thus the energy requirement for a temperature decrease remains dominated by the phononic heat capacity of Au, as used above. Nonetheless, one must consider the energy lost per electron; the average energy of an electron in a metallic system is approximately 3/5 of the Fermi energy ($\approx$5.5 eV for Au). Thus, for a 1 K temperature decrease at room temperature, approximately $7\times10^{8}$ electrons would need to be emitted from the Au surface.

To gain further insight into which of these mechanisms, or combination of mechanisms, is driving the observed cooling process, we perform two-temperature model (TTM) calculations \cite{Chen2006} for our experimental geometry. This method allows us to explicitly calculate energy losses/gains to the electron and lattice subsystems. While additional details can be found in the Supporting Information, we use the measured surface current as the heating source for the electronic subsystem, and the thermal properties (e.g., thermal conductivity and interfacial thermal resistances) of our 80 nm Au/Al\textsubscript{2}O\textsubscript{3} are determined from time-domain thermoreflectance (TDTR) measurements. To simulate the atomic or electronic ejection, we supply a `cooling' source that removes energy from either the phononic or electronic subsystem. While we ultimately find that either atom or electron ejections can re-produce our experimentally-measured data with high accuracy (see Fig.~2b), these calculations provide insight into the time-scale of this cooling process by accounting for heating at simultaneous time-scales; the best TTM fit to our data requires that the cooling occurs over the first 800 nanoseconds after the voltage is applied. 

We must now consider the temporal flux of various species incident upon the Au surface (Fig.~1) within these 800 nanoseconds.  To properly understand this, we compare the current measurements discussed above with time-resolved photoemission measurements; these experiments indicate that the only species present at the time-scales corresponding to the observed cooling are photons. Those results, shown in Fig.~3a, indicate that a fluence of photons impinges upon the Au surface at the time scales corresponding to the cooling and, importantly, they are present prior to the dramatic rise in measured current. The difference is characteristic of a pulsed plasma jet, produced with a remotely located active electrode. While the excited species that relax via photon emission and the charged particles are simultaneously produced near the electrode when the high voltage is applied, the photons arrive at the surface well before the comparatively low-velocity charged species. The wavelength range of photons emitted from the plasma jet, typical of plasma produced at atmospheric pressure mixtures of air and helium, extend from the IR to the extreme UV, with energies that range from below one eV to above 20 eV (see details in Supporting Information, including Figs.~S3-5). This range of photons incident to the gold surface are certainly sufficient to drive the emission processes discussed above. For gold, the work function is 5.1 eV and the enthalpy of atomization is 3.8 eV/atom (364 kJ mol$^{-1}$). Likewise, photon-driven desorption of adsorbates \cite{Zhou1991} has been observed for a wide range of material systems and photon energies. For example, the desorption of water from Pd(111) has been demonstrated using 6.4 eV (194 nm) and 5.0 eV (248 nm) photons \cite{Wolf1991}. The liberation of physisorbed species such as CO on Ag(111) \cite{Fleck1997} and H$_{2}$CO or CH$_{2}$CO on Ag(111) \cite{Howe1998} only require 1.1 eV (1098 nm) photons. While these species desorb as neutrals, oxygen can also be liberated from aluminum in the form of negative ions by photons in the range of about 8 to 10 eV \cite{Walton1998}. 

This observation of a photon-driven cooling mechanism is further supported by spatially-resolved thermoreflectance measurement results shown in Fig.~3b. For these data, we raster the plasma jet with respect to the position of the laser probe, allowing us to extract a spatio-temporal temperature profile of the Au surface. In agreement with our previous work \cite{Walton2018}, we observe a heating profile of $\approx0.5$ mm; this is approximately $1/3$ of the tube diameter from which the jet emanates from. More importantly, in this work, we observe cooling of the Au surface in regions extending beyond the heated width. This can be understood by recognizing the plasma is `guided' by the helium flow leaving the tube, such that the majority of the plasma-produced particles arrive at the surface over an area dictated by the diameter of the gas flow/surface intersection. Conversely, the photons formed via spontaneous emission from excited species formed along the plasma channel have both random phase and direction, thus potentially interacting with a much larger surface area.  The exception to this will be sub-200 nm photons, which are readily absorbed in the air outside the helium channel. Still, it is reasonable to expect a concentrated region of charged particle interactions toward the center and a more diffuse region of photon interactions extending beyond the region dominated by charged particles. 

While a photon-driven process is certain, both the liberation of surface adsorbates and electronic emission can be supported by our experimental measurements, two-temperature analysis, and prior TDTR measurements. Discerning the potential contribution of the two processes requires additional considerations.

The results of Fig.~3a clearly shows the photons arrive at the surface during the entire voltage pulse, as well as after the voltage is extinguished. That is, photons are present both before and after the charged particle flux that is associated with heating. If the cooling were caused by photon-driven electron emission, or ejection of gold atoms, one should expect cooling to occur during both times that photons are the dominate species since there is ample mass and essentially an infinite reservoir of electrons to be ejected. And yet, cooling is only observed before heating. On the other hand, photo-desorption of adsorbed species would only occur during the initial flux of photons, as the flux of all species - particularly heavy ions - will have liberated adsorbates from the surface prior to the end of the applied voltage pulse. Thus, photon-stimulated desorption of adsorbates is the most likely mechanism responsible for plasma cooling.

To further reinforce this plasma-induced evaporative cooling of a metal surface, we consider the relative heating and cooling contributions with varying plasma-jet parameters. For example, as the applied voltage of the electrodes is increased, we observe a corresponding increase in the measured surface current and the associated peak temperature of our thermoreflectance measurements (see Supporting Information and Fig.~S6); these observations can be attributed to an increase in the flux of charged species at the surface associated with a higher density plasma formed with increasing voltage. However, the minimum temperature achieved during cooling is negligibly affected by these changes, at least within the regimes considered in this work. Yet, like the flux of charged particles, the flux of photons responsible for either desorption or electronic emission will be increasing with voltage as well. In other words, the cooling effect has become saturated and the additional flux does not play a role. As noted previously, neither gold atoms nor electrons are in limited supply, and so atom- or electron mediated cooling mechanisms should not plateau. This further reinforces our posit that neither Au atom nor electron ejection is responsible for the observed plasma-induced cooling. A similar observation is observed with varying voltage pulse widths, whereby the net cooling magnitude is relatively constant above some critical width ($\approx2 \mu$s in this work), while the peak temperature varies significantly with changes in charged particle flux. Interestingly, below this critical pulse width, this cooling can occur $without$ any significant plasma-induced heating of the Au surface. Here, a sufficient number of high-energy photons are produced at plasma ignition to drive desorption, but the voltage is extinguished before the streamer and associated flux of charged particles can reach the surface with sufficient intensity. Just above this threshold, a spike in both the heating and cooling profile is observed due to an abundance of both charged particles and photons. These observations not only reinforce the notion of photo-driven, desorption-induced cooling of the Au surface, but also indicate potential regimes for which the relative heating and cooling profiles of the target surface can be manipulated through various plasma parameters. In fact, our previous work \cite{Walton2018} suggested conditions where net cooling was a possibility, but in the absence of an explanation, it remained a curiousity.

\paragraph*{Summary.} In summary, we have provided the first nanosecond-resolved measurement of energy transduction during plasma-surface interactions with a transient thermoreflectance method. Simple ohmic heating of the metal Au surface through a flux of charge species is the predominant mechanism of energy transfer, which dissipates primarily through conduction into the material; this process is in excellent agreement with our two-temperature model calculations. Furthermore, we find that the plasma jet induces transient cooling of the Au surface prior to Joule heating. Over the course of approximately 800 nanoseconds, the plasma jet liberates surface species from the Au film, thus temporarily cooling the metal surface. This unique process, typically overwhelmed by the resistive heating of material surfaces during steady-state operation and thus undetectable, opens the door to a new means of surface cooling and provides critical insight to the plasma-material interactions that drive material modification and processing.

\paragraph*{Acknowledgements.} This manuscript is based upon work supported in part by the Office of Naval Research under award No. N00014-20-1-2686 and by the Air Force Office of Scientific Research under award no. FA9550-18-1-0352. This work was also partially supported by the Naval Research Laboratory base program.

\bibliography{PlasmaCooling}
\bibliographystyle{Science}

 \begin{figure}
	\centering
	\includegraphics[width=1\columnwidth]{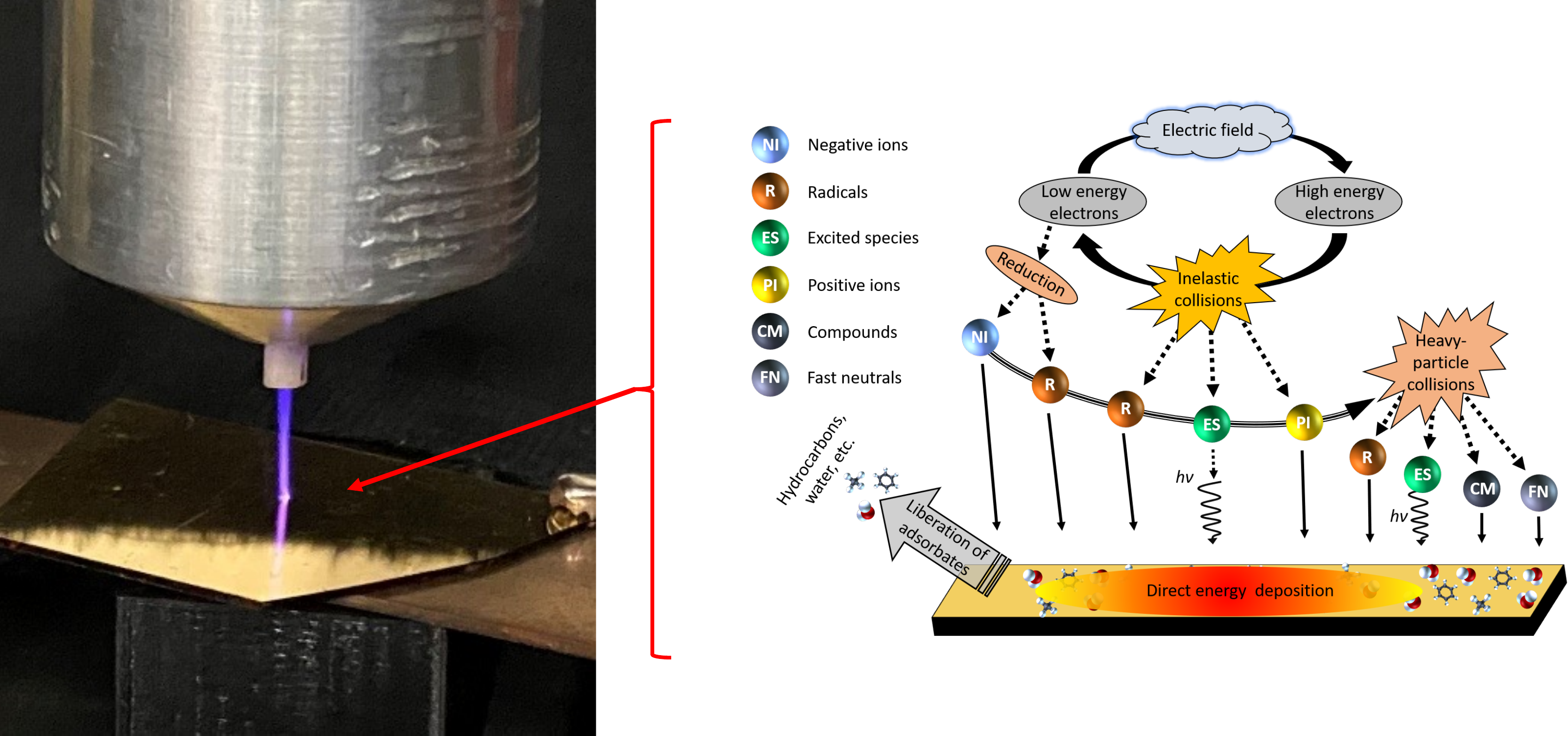}
	\caption{Left.) Photograph of the atmospheric plasma jet interacting with a thin Au film on sapphire substrate. Right.) Cartoon schematic of the various species and physical processes and resulting species produced within the jet along with their respective interactions with the metal film.} 
	\label{Figure1}
\end{figure}

 \begin{figure}
	\centering
	\includegraphics[width=1\columnwidth]{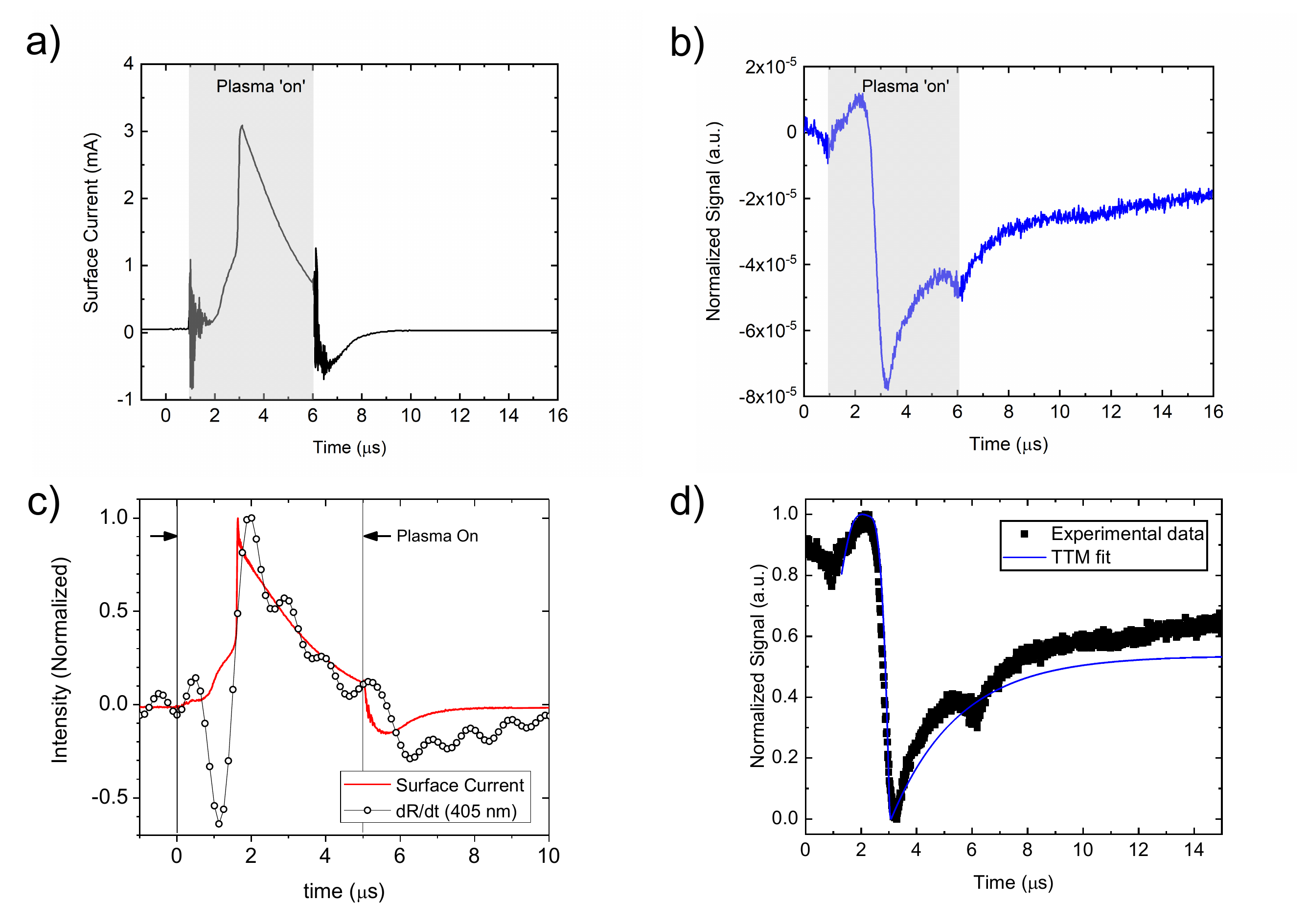}
	\caption{a) The measured surface current of the Au film and b) the measured thermoreflectance of the Au surface as a function of time for a 5 $\mu$s plasma pulse. The laser probe wavelength is 654 nm. c) Measured surface current (solid red line) overlaid with the temporal derivative of the measured changed in reflectivity due to the plasma pulse (open circles). The laser probe wavelength here is 405 nm. The two are in reasonable agreement, aside from the reduction in surface temperature around 1 $\mu$s, indicating Joule heating to be the primary mechanism of energy transfer from the plasma to the Au surface. d) Measured thermoreflectance data from (b) due to the plasma pulse (black dots) and two-temperature model calculations for the sample system (solid blue line). The observed cooling is attributed to the removal of adsorbed species on the Au surface.}
	\label{Figure2}
\end{figure}

\begin{figure}
	\centering
	\includegraphics[width=1\columnwidth]{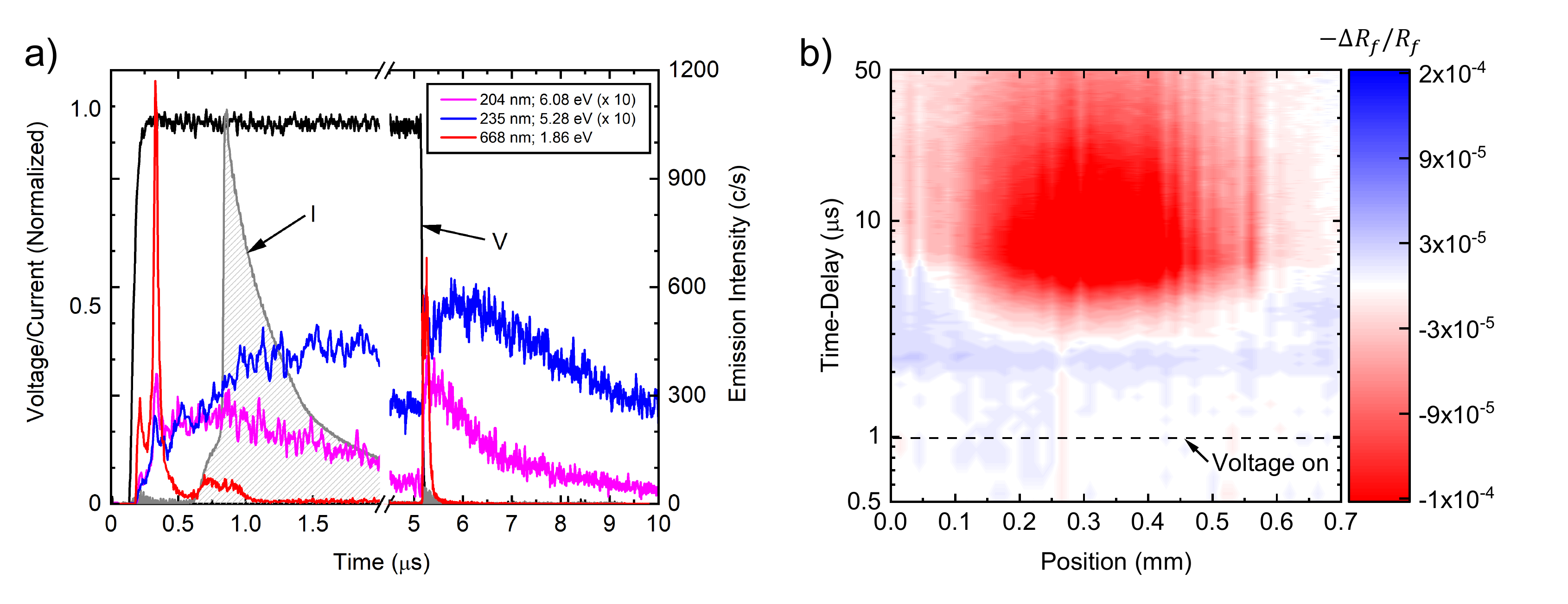}
	\caption{a) Time-resolved emission measurements of the He jet interacting with gold surface along with the voltage applied to the electrode and the measured surface current. The emission lines shown are from He$^{\ast}$ at 668 nm and NO$^{\ast}$ at 204 nm and 235 nm, illustrating that a range of energetic photons arrive at the surface both before and after the charged particle flux. b) Spatially-resolved thermoreflectance measurements of an Au surface. A width of 0.5 mm is heated from the flux of charged particles, while the observed photon-induced cooling extends to a much larger region. Blue corresponds to observed cooling of the sample surface (increase in $\Delta R_{f}$), while red denotes an increase in surface temperature (decrease in $\Delta R_{f}$).}
	\label{Figure4}
\end{figure}

\end{document}